\documentclass[conference,letterpaper]{IEEEtran}

\usepackage{balance}

\usepackage[utf8]{inputenc} 
\usepackage[T1]{fontenc}
\usepackage{ifthen}

\usepackage{tikz}
\newcommand\copyrighttext{%
    \footnotesize \textcopyright\ 2026 IEEE. Personal use of this material is permitted. 
    Permission from IEEE must be obtained for all other uses, in any current or future media, 
    including reprinting/republishing this material for advertising or promotional purposes, 
    creating new collective works, for resale or redistribution to servers or lists, or reuse 
    of any copyrighted component of this work in other works.}
\newcommand\copyrightnotice{%
    \begin{tikzpicture}[remember picture,overlay]
        \node[anchor=north,yshift=-10pt] at (current page.north) 
            {\fbox{\parbox{\dimexpr\textwidth-\fboxsep-\fboxrule\relax}{\copyrighttext}}};
    \end{tikzpicture}%
}

\usepackage[cmex10]{amsmath} 
\usepackage{amsfonts}
\usepackage{amssymb}         

\usepackage{stfloats}
\usepackage{graphicx}
\usepackage{epstopdf}
\usepackage[caption=false,subrefformat=parens,labelfont=rm,textfont=rm]{subfig}

\usepackage{color}
\usepackage{xcolor}
\usepackage{url}
\usepackage{soul}
\DeclareRobustCommand{\mhl}[1]{%
	\ifmmode\text{\rvp{$#1$}}\else\revisedpart{#1}\fi
}

\newcommand{\rvp}[1]{{\color{blue}{#1}}}

\usepackage{cite}
\makeatletter 
\let\NAT@parse\undefined
\makeatother
\usepackage[colorlinks,citecolor=blue]{hyperref}
\PassOptionsToPackage{unicode}{hyperref}
\PassOptionsToPackage{naturalnames}{hyperref}
\hypersetup{
    urlcolor=blue,                
    linkcolor= blue,                
    bookmarks=false,                 
    bookmarksopen=false,
    filecolor=black,
    citecolor=blue,
    linkbordercolor=blue
}

\usepackage[nameinlink,noabbrev]{cleveref}

\crefname{equation}{}{}
\crefname{figure}{Fig.}{Fig.}
\crefname{table}{Table}{Table}
\crefname{lemma}{Lemma}{Lemma}
\crefname{prop}{Proposition}{Proposition}
\crefname{thm}{Theorem}{Theorem}
\crefname{corol}{Corollary}{Corollary}
\crefname{defn}{Definition}{Definition}
\crefname{rem}{Remark}{Remark}
\crefname{insight}{Insight}{Insight}
\Crefname{algocf}{Algorithm}{Algorithm}
\Crefname{appendix}{Appendix}{Appendix}

\usepackage{amsthm}

\usepackage{booktabs}
\usepackage{multicol}
\usepackage{threeparttable}
\usepackage{multirow}
\usepackage{tabularx}
\newcolumntype{L}[1]{>{\raggedright\arraybackslash}p{#1}}
\newcolumntype{C}[1]{>{\centering\arraybackslash}p{#1}}
\newcolumntype{R}[1]{>{\raggedleft\arraybackslash}p{#1}}

\usepackage[linesnumbered,ruled,vlined]{algorithm2e}
\usepackage{setspace}
\SetAlgoSkip{}
\SetAlgoInsideSkip{}
\SetAlFnt{\small}
\SetAlCapFnt{\small}
\SetAlCapNameFnt{\small}
\SetKwRepeat{Do}{do}{while}
\SetEndCharOfAlgoLine{}
\SetAlgoCaptionSeparator{}

\begin{document}
\title{User-Centric Clustering for uRLLC in Cell-Free RAN via Extreme Value Theory} 


\author{%
\IEEEauthorblockN{Yu Zhang\textsuperscript{1}, Xinyue Yang\textsuperscript{1}, Dongming Wang\textsuperscript{2,3}, Boyou Yi\textsuperscript{1}, Yaqin Xie\textsuperscript{1}, Hua Zhou\textsuperscript{1},  and Zhizhong Zhang\textsuperscript{1}}
  \IEEEauthorblockA{
  \textsuperscript{1}School of Electronics and Information Engineering, 
                    Nanjing University of Information Science and Technology, \\
                    Nanjing 210044, China\\
\textsuperscript{2}National Mobile Communications Research Laboratory, 
                    Southeast University, Nanjing 210096, China\\
\textsuperscript{3}Purple Mountain Laboratories, Nanjing 211111, China\\
                    Email: zhangyu@nuist.edu.cn}
}

\maketitle

\copyrightnotice

\begingroup
\renewcommand{\thefootnote}{} 
\endgroup

\begin{abstract}
Ultra-reliable low-latency communication (uRLLC) is a pivotal enabler for B5G/6G networks, yet it faces severe challenges from rare but critical extreme events, which are characterized by heavy tails in the delay distribution. 
While the cell-free radio access network (CF-RAN) architecture offers essential spatial diversity to combat these uncertainties, conventional user-centric clustering designs typically focus on average metrics, thereby inadequately addressing such tail behaviors. 
We propose a novel, tail-risk-aware, user-centric clustering framework operating within the finite blocklength (FBL) regime.
Our approach employs extreme value theory (EVT), specifically the peaks-over-threshold (POT) model, to accurately quantify the probability of queue latency violations. 
This framework is applied to formulate an energy efficiency (EE) maximization problem under strict tail latency constraints.
The problem is solved via an efficient online algorithm that integrates Lyapunov optimization with successive convex approximation (SCA). 
Simulation results demonstrate that the proposed scheme, through its dynamic adaptation of cluster formation to mitigate tail risks, achieves a superior reliability-efficiency trade-off and leads to a significant suppression of extreme latency events.


\end{abstract}

\section{Introduction}
\label{sec:introduction}
Ultra-reliable low-latency communication (uRLLC) is a key capability of future wireless networks for mission-critical applications such as factory automation and remote control, where stringent latency and reliability requirements must be satisfied simultaneously~\cite{zhang2025performance}. 
One promising implementation path is provided by cell-free massive multiple-input multiple-output (CF-mMIMO) and its deployment-efficient cell-free radio access network (CF-RAN) architecture, which leverages cooperation among distributed access points (APs) to enhance macro-diversity and coverage~\cite{ren2024tight,wang2023full,gottsch2023user}. 
Furthermore, scalable CF-RAN architectures facilitate network coordination and control while maintaining manageable signaling overhead~\cite{cao2025implementation}.

To reduce end-to-end latency in uRLLC, short-packet transmission in the finite blocklength (FBL) regime is widely adopted. 
However, this approach introduces non-negligible decoding errors and a corresponding channel-dispersion penalty to the achievable rate, which complicates resource allocation~\cite{peng2023resource}.
User-centric clustering offers a robust solution to these issues, serving as a key enabler for CF-RAN.  
This strategy selectively associates APs with user equipments (UEs) to preserve macro-diversity and cooperation gains, while simultaneously maintaining manageable signaling overhead~\cite{interdonato2020local,guo2025stochastic}. 
When integrated with fairness scheduling, user-centric clustering can further enhance interference mitigation under stringent uRLLC requirements~\cite{gottsch2023user,wang2024uc_urllc}.
Typical problem formulations involve max rate optimization with binary AP-UE association variables, which are often solved using heuristic methods due to practical combinatorial constraints~\cite{wang2024uc_urllc}.
For long-term operation, clustering strategies can be integrated with Lyapunov optimization to meet statistical quality-of-service (QoS) targets~\cite{chong2024statistical}.
Moreover, joint designs frequently incorporate AP activation coupled with sleep control mechanisms to enhance energy efficiency (EE) for uRLLC~\cite{yan2025efficient}.

A critical challenge is that severe performance degradations in uRLLC often stem from low-probability but high-severity congestion spikes.
Such tail risks are difficult to capture and suppress using only average EE objectives or low-order statistical QoS constraints~\cite{chong2024statistical}. 
Specifically, permanently maintaining large clusters to guard against rare spikes can severely degrade EE, while failing to react promptly leaves the system vulnerable to service outages.
To address this gap, we introduce extreme value theory (EVT), a powerful statistical framework for modeling rare events, to explicitly characterize and constrain the tail behavior of queue backlogs~\cite{perez2024extreme}.

Motivated by these challenges, we propose an EVT-driven user-centric clustering framework. 
Specifically, we leverage EVT to model the tail distribution of queue backlogs using the peaks-over-threshold (POT) method. By deriving explicit tail-risk indicators from the generalized Pareto distribution (GPD), we transform the elusive extreme events into tractable constraints. 
This allows adaptive AP-UE association reconfiguration when the queue state triggers tail risks, thereby suppressing extreme queueing events. 
Formulating this tail-risk-aware clustering mathematically leads to a non-convex problem that is challenging to solve in real-time. 
To provide a practical online solution, we develop tractable tail-risk measures and propose a centralized per-slot algorithm based on iterative successive convex approximation (SCA) with low complexity. 
Simulation results demonstrate that the proposed scheme achieves a superior reliability-efficiency trade-off, significantly reducing queue violations while maintaining competitive EE against representative baselines.

\begin{figure}[t]
  \centering
  \includegraphics[width=0.95\columnwidth]{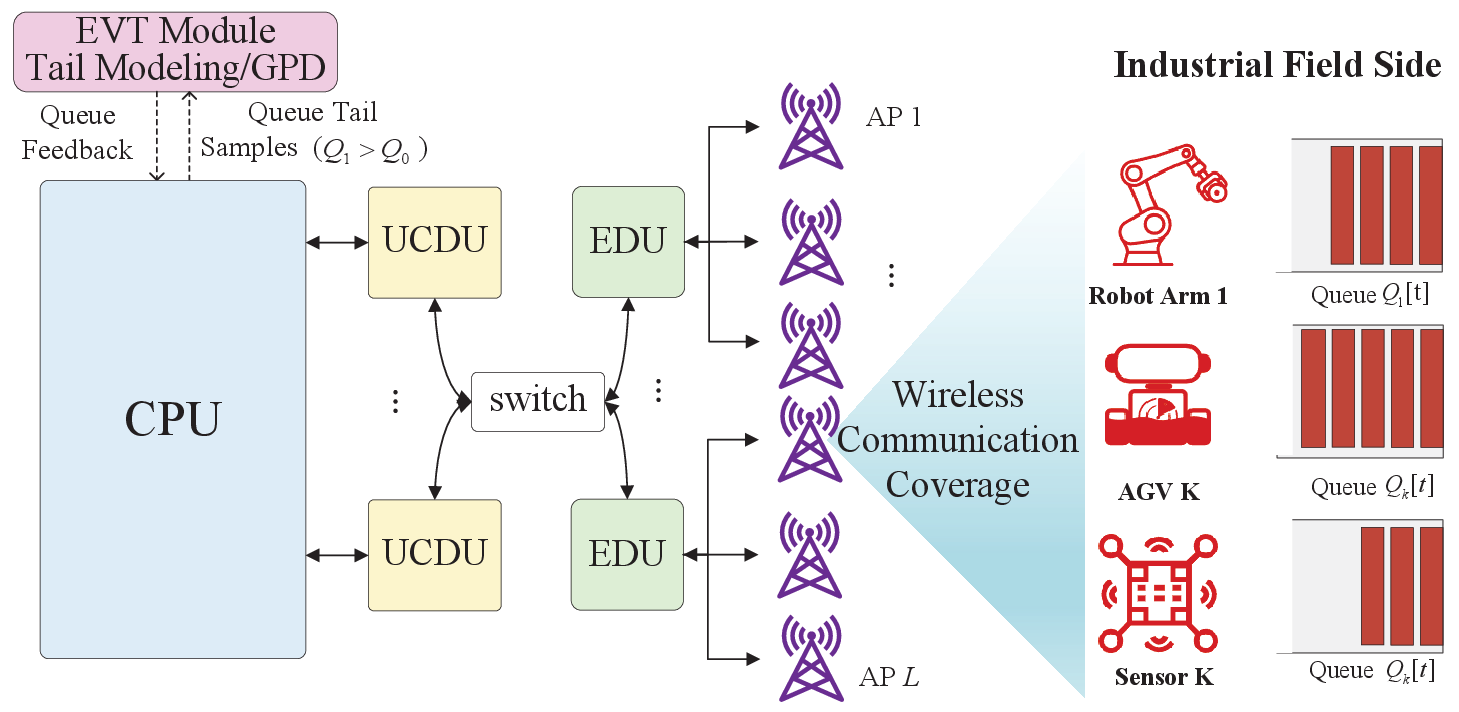}
  \caption{Illustration of the CF-RAN system with EVT-aware clustering.}
  \label{fig:arch_evt}
\end{figure}

\section{System Model}
\label{sec:system_model}
\addtolength{\topmargin}{0.25in} 

\subsection{Channel Model}\label{subsec:channel_model}
We consider a CF-RAN architecture operating in time-division duplex mode, as illustrated in Fig.~\ref{fig:arch_evt}.
The system comprises $M$ edge distributed units (EDUs) and $L$ distributed APs equipped with $N$ antennas to serve $K$ single-antenna UEs.
In addition, user-centered distributed units (UCDUs) act as logical coordination entities that aggregate UE-specific state information and assist the central processing unit (CPU) in user-centric clustering updates with reduced signaling overhead\cite{guo2025stochastic}.
A coherence block contains $\tau_{\rm{c}}$ channel uses, where $\tau_{\rm{p}}$ is reserved for uplink (UL) pilots.

The channel vector between UE $k$ and AP $l$ associated with EDU $m$ is modeled as $\mathbf{g}_{k,l,m}[t]=\sqrt{\beta_{k,l,m}}\,\mathbf{h}_{k,l,m}[t]$, where $\beta_{k,l,m}$ represents the large-scale fading coefficient, encompassing path loss and shadowing, which is assumed to be quasi-static. 
The small-scale fading component follows a Rayleigh fading model, distributed as  $\mathbf{h}_{k,l,m}\sim\mathcal{CN}(\mathbf{0},\mathbf{I}_N)$.

During each coherence block, UEs transmit UL pilots of length $\tau_{\rm{p}}$.
Since $\tau_{\rm{p}}<K$, pilot reuse is inevitable, leading to pilot contamination.
For simplicity, we denote AP $l$ associated with EDU $m$ as AP $(l,m)$.
Via minimum-mean-square-error (MMSE) estimation~\cite{demir2021foundations}, the estimated channel and estimation error for UE $k$ at AP $(l,m)$ are defined as $\hat{\mathbf{g}}_{k,l,m}[t]$ and $\tilde{\mathbf{g}}_{k,l,m}[t]\triangleq \mathbf{g}_{k,l,m}[t]-\hat{\mathbf{g}}_{k,l,m}[t]$, respectively. Their distributions are  $\hat{\mathbf{g}}_{k,l,m}[t] \sim \mathcal{CN}(\mathbf{0},\gamma_{k,l,m}\mathbf{I}_N)$ and $\tilde{\mathbf{g}}_{k,l,m}[t] \sim \mathcal{CN}(\mathbf{0},(\beta_{k,l,m}-\gamma_{k,l,m})\mathbf{I}_N)$,
where
$\gamma_{k,l,m}\triangleq \frac{\tau_{\rm{p}} \rho_{\rm{p}} \beta_{k,l,m}^2}{\tau_{\rm{p}} \rho_{\rm{p}} \sum_{i\in\mathcal{P}_k}\beta_{i,l,m}+1}$,
with $\rho_{\rm{p}}$ being the normalized pilot power per symbol and $\mathcal{P}_k$ denoting the pilot-sharing set of UE $k$.

\subsection{Downlink Transmission}\label{subsec:downlink_transmission}

We consider DL transmission in slot $t$ with user-centric clustering modeled by binary association variables
\begin{align}
\text{C1:}\;
    a_{k,l,m}[t]\in\{0,1\}.
\end{align}
Define the serving-AP set of UE $k$ as
$\mathcal{M}_k[t]\triangleq\{(l,m)\,|\,a_{k,l,m}[t]=1\}$,
and the served-UE set of AP $(l,m)$ as
$\mathcal{K}_{l,m}[t]\triangleq\{k\,|\,a_{k,l,m}[t]=1\}$.
The DL transmit signal of AP $(l,m)$ is
\begin{align}
\mathbf{x}_{l,m}[t]
=
\sum\limits_{k=1}^{K}
\sqrt{\rho_{\mathrm d}\,\bar{\eta}_{k,l,m}[t]}\,
\mathbf{w}_{k,l,m}[t]\,
s_k[t],
\label{eq:dl_tx_signal}
\end{align}
where $s_k[t]$ is the unit-power data symbol, $\rho_{\mathrm d}$ is the normalized DL transmit power, and $\bar{\eta}_{k,l,m}[t]\triangleq \eta_{k,l,m}[t]\,a_{k,l,m}[t]$.
For simplicity, we assume equal power allocation
$\eta_{k,l,m}[t]=1/|\mathcal{K}_{l,m}[t]|$ for $k\in\mathcal{K}_{l,m}[t]$, and $0$ otherwise.

Adopting the localized full-pilot zero-forcing (FZF) precoding scheme~\cite{interdonato2020local}, each AP $(l,m)$ constructs a local matrix spanning the $\tau_{\rm{p}}$-dimensional pilot subspace:
\begin{align}
\hat{\mathbf G}_{l,m}[t]
\triangleq
\big[\hat{\mathbf g}_{l,m,1}^{\mathrm p}[t],\ldots,\hat{\mathbf g}_{l,m,\tau_{\rm{p}}}^{\mathrm p}[t]\big]
\in\mathbb{C}^{N\times \tau_{\rm{p}}},
\label{eq:Ghat_pilot_subspace}
\end{align}
where $\hat{\mathbf g}_{l,m,i}^{\mathrm p}[t]$ is obtained from correlating the received UL pilot signal with pilot $\boldsymbol{\phi}_i$, and $i_k\in\{1,\ldots,\tau_{\rm{p}}\}$ denotes the pilot index assigned to UE $k$.
Under the typical assumption that $N > \tau_{\rm{p}}$, the local FZF precoder at AP $(l,m)$ for serving UE $k$ is given by
\begin{align}
\mathbf w_{k,l,m}[t]
=
\frac{
\hat{\mathbf G}_{l,m}[t]\big(\hat{\mathbf G}_{l,m}^{\mathrm H}[t]\hat{\mathbf G}_{l,m}[t]\big)^{-1}\mathbf e_{i_k}
}{
\sqrt{\mathbb{E}\!\left\{\left\|
\hat{\mathbf G}_{l,m}[t]\big(\hat{\mathbf G}_{l,m}^{\mathrm H}[t]\hat{\mathbf G}_{l,m}[t]\big)^{-1}\mathbf e_{i_k}
\right\|^2\right\}}}
,
\label{eq:fpzf_precoder_one_line}
\end{align}
where $\mathbf e_{i_k}$ is the $i_k$-th standard basis vector.
Applying the use-and-then-forget (UatF) bounding technique and the standard closed-form analysis for FZF precoding yields the effective signal-to-interference-plus-noise ratio (SINR) in~\eqref{eq:gammahat_fzf_closed_form}, shown at the bottom of the next page.
\begin{figure*}[b]
\rule{\linewidth}{0.4pt}
\begin{align}
\hat{\gamma}_{k}^{\mathrm{FZF}}[t]
=
\frac{
\rho_{\mathrm d}(N-\tau_{\rm p})
\Big(\sum\limits_{(l,m)\in\mathcal M_k[t]}
\sqrt{\bar{\eta}_{k,l,m}[t]\;\gamma_{k,l,m}[t]}\Big)^2
}{
\rho_{\mathrm d}
\sum\limits_{i=1}^{K}
\sum\limits_{(l,m)\in\mathcal M_i[t]}
\bar{\eta}_{i,l,m}[t]\big(\beta_{l,m,k}[t]-\gamma_{k,l,m}[t]\big)
+
\rho_{\mathrm d}(N-\tau_{\rm p})
\sum\limits_{i\in\mathcal P_k\setminus\{k\}}
\Big(\sum\limits_{(l,m)\in\mathcal M_i[t]}
\sqrt{\bar{\eta}_{i,l,m}[t]\;\beta_{k,l,m}[t]}\Big)^2
+1
}.
\label{eq:gammahat_fzf_closed_form}
\end{align}
\end{figure*}

To characterize the performance in the FBL regime pertinent to uRLLC, we employ the normal approximation for FBL channels along with a Jensen-type inequality to derive a tractable closed-form lower bound, following the approaches in~\cite{peng2023resource,polyanskiy2010finite}. 
Denote by $\tau_{\rm{c}}$ the channel coherence block length. 
The pilot fraction is defined as $\eta_{\rm p} \triangleq \tau_{\rm{p}} / \tau_{\rm{c}}$.
Then, a closed-form lower bound of the achievable DL spectral efficiency of UE $k$ in slot $t$ is expressed as
\begin{align}
\hat R_k[t]
=
\frac{1-\eta_{\rm p}}{\ln 2}\,
f_k\!\left(\frac{1}{\hat\gamma_{k}^{\mathrm{FZF}}[t]}\right),
\label{eq:fbl_rate_closed_form}
\end{align}
where $\varepsilon_k$ is the target decoding error probability, $Q^{-1}(\cdot)$ is the inverse Gaussian $Q$-function, and 
\begin{align}
f_k(x)\triangleq
\ln\!\left(1+x\right)-
\frac{Q^{-1}(\varepsilon_k)}{\sqrt{\tau_{\rm{c}}(1-\eta_{\rm p})}}
\sqrt{\frac{2x+1}{(1+x)^2}}.
\label{eq:fk_def}
\end{align}

\subsection{Queue Model}
\label{subsec:queue_model}

The system operates in discrete time slots indexed by $t=0,1,2,\cdots$, with each slot spanning one channel coherence block. 
Each UE $k$ maintains a data buffer with its backlog (in bits) at the end of slot $t$ denoted by $Q_k[t]$. 
In each slot $t$,  UE $k$ experiences an exogenous data arrival of $A_k[t]$ and receives a service of $\hat{R}_k[t]$.
Consequently, the queue backlog evolves as $Q_k[t+1]=\left[Q_k[t]-\hat{R}_k[t]\right]^+ + A_k[t]$, where $[x]^+\triangleq \max\{x,0\}$.
The exogenous arrivals are assumed to be uniformly bounded for all users and time slots:
\begin{align}
\text{C2:}\; 0\le A_k[t]\le A_k^{\max},\; \forall k,t .
\label{cons:C5_arrival}
\end{align}

To quantify congestion and characterize rare events, we introduce an exceedance threshold $Q_0$.
A tail event for UE $k$ in slot $t$ is then defined as the occurrence of $Q_k[t]\ge Q_0$.
To control the frequency of such tail events, a long-term exceedance probability constraint is imposed on each UE:
\begin{align}
\text{C3:}\; \lim_{T\to\infty}\frac{1}{T}\sum\nolimits_{t=0}^{T-1}\Pr\!\left(Q_k[t]\ge Q_0\right)
\le \varepsilon_k^{(Q)},\; \forall k .
\label{cons:C6_overflow}
\end{align}
To further capture the severity of tail events, beyond their mere frequency, we define the excess queue length beyond the threshold as $Z_k[t]\triangleq \bigl[Q_k[t]-Q_0\bigr]^+$.
Building on the excess process, the time-average first and second moments are defined as $\bar Z_k^{(1)}$ and $\bar Z_k^{(2)}$, respectively.
We then place constraints on these moments to limit the long-term average severity of exceedances:
\begin{subequations}
  \begin{align}
\text{C4:}\; \bar Z_k^{(1)} &\triangleq \lim_{T\to\infty}\frac{1}{T}\sum\nolimits_{t=0}^{T-1}\mathbb{E}\!\left[Z_k[t]\right]
\le \zeta_k^{(1)},\; \forall k ,
\label{cons:C7_tail_m1}
\\
\text{C5:}\; \bar Z_k^{(2)}  &\triangleq \lim_{T\to\infty}\frac{1}{T}\sum\nolimits_{t=0}^{T-1}\mathbb{E}\!\left[Z_k^2[t]\right]
\le \zeta_k^{(2)},\;\forall k .
\label{cons:C8_tail_m2}
\end{align}  
\end{subequations}
The parameters $(\varepsilon_k^{(Q)},\zeta_k^{(1)},\zeta_k^{(2)})$ in C3–C5 provide a comprehensive means of jointly controlling the frequency and the severity of rare exceedances above the threshold $Q_0$.

\subsection{Power Consumption Model}
The total power consumption is modeled following the approach in ~\cite{ngo2018total_ee_cf}.
Let $P_{l,m}[t]$ denote the power consumption at AP~$(l,m)$ in slot $t$.
This consumption comprises three components:
\begin{align}
  P_{l,m}[t]
  &=
  \frac{\rho_{\mathrm d}}{\alpha_m}
  \sum\nolimits_{k=1}^{K}
    \eta_{k,l,m}[t]\,a_{k,l,m}[t] \nonumber\\
   & \quad  + P_{\mathrm{link}}
  \sum\nolimits_{k=1}^{K}
    a_{k,l,m}[t] 
+ P_{\mathrm{cir}},
  \label{eq:Plm_def}
\end{align}
where $\alpha_m\in(0,1]$ is the efficiency of power amplifiers, $P_{\mathrm{cir}}$ represents the traffic circuit power, and $P_{\mathrm{link}}$ is the per-link load-dependent power coefficient.

The power consumption on the EDU side is modeled as a constant $P_{\mathrm{edu}}$~\cite{chen2023energy}.
Consequently, the total network power consumption in time slot $t$ is given by
\begin{align}
P_{\mathrm{tot}}[t]
\triangleq
\sum\nolimits_{m=1}^{M}\sum\nolimits_{l=1}^{L} P_{l,m}[t] + P_{\mathrm{edu}}.
\label{eq:Ptot_def}
\end{align}
To capture the trade-off between long-term EE and network stability, the long-term system EE is defined as~\cite{song2025joint}
\begin{align}
\overline{\mathrm{EE}}
&\triangleq
\frac{\lim\limits_{T\to\infty}\frac{1}{T}\sum\nolimits_{t=0}^{T-1}\mathbb{E}\!\left\{B\,R_{\mathrm{sum}}[t]\right\}}
{\lim\limits_{T\to\infty}\frac{1}{T}\sum\nolimits_{t=0}^{T-1}\mathbb{E}\!\left\{P_{\mathrm{tot}}[t]\right\}}
=
\frac{B\,\bar R_{\mathrm{sum}}}{\bar P_{\mathrm{tot}}}.
\label{eq:EE_def}
\end{align}
where $B$ denotes the system bandwidth, and $R_{\mathrm{sum}}[t]=\sum\nolimits_{k=1}^{K}\hat R_k[t]$ defines the sum rate.
The numerator and denominator of $\bar{\mathrm{EE}}$ correspond to the time-averaged sum throughput $B\bar{R}_{\mathrm{sum}}$ and the time-averaged total power consumption $\bar{P}_{\mathrm{tot}}$, respectively.

\subsection{Problem Formulation}\label{sec:problem_formulation}
Let $\mathbf{X}[t]\triangleq \{a_{k,l,m}[t],\,\forall k,l,m\}$ denote the collection of all AP--UE association variables in slot $t$.
This paper addresses the problem of maximizing the long-term system EE by optimizing the user-centric clustering decisions:
\begin{align}
 \textbf{P1}:\ \max_{\{\mathbf{X}[t]\}} \ \overline{\mathrm{EE}}
\quad \text{s.t. C1-C5}.
  \label{eq:prob_obj}
\end{align}
Problem \textbf{P1} is an infinite-horizon stochastic optimization problem with binary decision variables and long-term constraints, which is large-scale and inherently non-convex.
To tackle this problem, we leverage a Lyapunov drift-plus-penalty framework to decompose the long-term stochastic optimization into a sequence of per-slot deterministic problems, thereby enabling the design of a low-complexity online policy that approximately solves~\textbf{P1}.

\section{Proposed User-Centric Clustering Method}
In this section, we propose an online solution to the long-term EE maximization problem \textbf{P1} using Lyapunov optimization.
Specifically, we consider an EVT-aware dynamic clustering that adapts AP-UE associations to queue-tail states.

\subsection{EVT-Based Queue Tail Modeling}\label{subsec:evt_tail}
This subsection adopts EVT to model the tail behavior of queue exceedances and to support the long-term tail constraints C3-C5 in \textbf{P1}.
The resulting tail descriptors enable joint control of the exceedance frequency and severity.
We define the conditional CDF
\begin{align}
G_k(z)\triangleq \Pr\!\left(Z_k[t]\le z \mid Z_k[t]>0\right).
\end{align}
Under mild regularity conditions, the Pickands-Balkema-de Haan theorem states that, when the threshold $Q_0$ is sufficiently high and the observation horizon is long,
the conditional exceedance distribution $G_k(z)$ is well approximated by a GPD with shape parameter $\xi_k$ and scale parameter $\sigma_k>0$:
\begin{align}
G_k(z)
=
1-\left(1+\xi_k z/{\sigma_k}\right)^{-1/\xi_k},
\quad z \ge 0,
\label{eq:gpd_def}
\end{align}
where $1+\xi_k z/\sigma_k>0$, with the usual continuous limit taken as $\xi_k\to 0$.
Provided $\xi_k<1/2$, the first two conditional moments exist and admit the standard closed forms
\begin{subequations}\label{eq:gpd_moments}
\begin{align}
\mathbb{E}\!\left[Z_k \mid Z_k>0\right]
&=
{\sigma_k}/\left({1-\xi_k} \right),
\label{eq:gpd_moments_a}\\
\displaybreak[2]
\mathbb{E}\!\left[Z_k^2 \mid Z_k>0\right]
&=
{2\sigma_k^2} / \left( 1-3\xi_k+2\xi_k^2\right).
\label{eq:gpd_moments_b}
\end{align}
\end{subequations}

In practice, under a steady-state regime, queue samples $\{Q_k[t]\}$ are collected over an extended time horizon, from which the positive exceedances $\{Z_k[t]:Z_k[t]>0\}$ are extracted.
We then estimate the GPD parameters $(\xi_k,\sigma_k)$ from these samples via maximum likelihood estimation (MLE).
These parameters serve as interpretable tail descriptors: $\xi_k$ captures tail heaviness, while $\sigma_k$ governs the scale of exceedances.
Since $Z_k[t]=0$ whenever $Q_k[t]<Q_0$, the law of total expectation yields, for each slot $t$,
\begin{subequations}\label{eq:moment_factorization}
\begin{align}
\mathbb{E}\!\left[Z_k[t]\right]
&=
p_k^{\mathrm{exc}}[t]\,
\mathbb{E}\!\left[Z_k[t]\mid Z_k[t]>0\right],
\label{eq:moment_factorization_a}\\
\mathbb{E}\!\left[Z_k^2[t]\right]
&=
p_k^{\mathrm{exc}}[t]\,
\mathbb{E}\!\left[Z_k^2[t]\mid Z_k[t]>0\right].
\label{eq:moment_factorization_b}
\end{align}
\end{subequations}
Substituting \eqref{eq:moment_factorization} into C4 and C5 reveals that the tail-moment constraints jointly depend on the exceedance frequency $p_k^{\mathrm{exc}}[t]$ and the conditional exceedance severity (conditional moments).
Under the POT and GPD approximation, the conditional moments admit the closed forms in \eqref{eq:gpd_moments}, yielding a computable tail descriptor in terms of $(p_k^{\mathrm{exc}}[t],\xi_k,\sigma_k)$.

\subsection{Lyapunov Framework and Virtual Queues}
Building upon the EVT-based tail characterization, we next integrate the average queue length and tail moment constraints into the design of an online control policy via Lyapunov optimization. 
To enforce the long-term constraints in an online manner, we introduce, for each UE $k$, three virtual queues alongside the physical queue $Q_k[t]$: (i) $U_k[t]$ for the overflow-probability constraint
  $\Pr(Q_k[t]\ge Q_0)\le \varepsilon_k^{(Q)}$; (ii) $Y_k[t]$ for the first tail-moment constraint \eqref{cons:C7_tail_m1}; (iii) $W_k[t]$ for the second tail-moment constraint \eqref{cons:C8_tail_m2}.
Their updates are given by
\begin{subequations}\label{eq:virtual_queues}
\begin{align}
U_k[t+1]
&=
\left[
  U_k[t] + \mathbf{1}_{\{Q_k[t]\ge Q_0\}} - \varepsilon_k^{(Q)}
\right]^+,
\label{eq:virtual_queues_a}\\
Y_k[t+1]
&=
\left[
  Y_k[t] + Z_k[t] - \zeta_k^{(1)}
\right]^+,
\label{eq:virtual_queues_b}\\
W_k[t+1]
&=
\left[
  W_k[t] + Z_k^2[t] - \zeta_k^{(2)}
\right]^+.
\label{eq:virtual_queues_c}
\end{align}
\end{subequations}
Define the concatenated queue state vector as $  \boldsymbol{\Theta}[t] \triangleq \left\{ Q_k[t], U_k[t], Y_k[t], W_k[t] : k=1,\cdots,K \right\}$, which aggregates all physical and virtual queues.
We define the quadratic Lyapunov function
\begin{align}
  L(\boldsymbol{\Theta}[t])
  =
  \frac{1}{2}
  \sum_{k=1}^K
  \bigl(Q_k^2[t]
    + U_k^2[t]
    + Y_k^2[t]
    + W_k^2[t]
  \bigr),
\end{align}
and define the one-slot conditional Lyapunov drift as
$\Delta[t]
=\mathbb{E}\left[
  L(\boldsymbol{\Theta}[t+1]) - L(\boldsymbol{\Theta}[t])
  \,\middle|\,
  \boldsymbol{\Theta}[t]
\right]$.
Denote by $\mathrm{EE}[t] = R_{\mathrm{sum}}[t] / P_{\mathrm{tot}}[t]$ the instantaneous EE in time slot $t$. 
The drift-plus-penalty term is defined as 
\begin{align}
 \Delta_V[t]\triangleq \Delta[t]-V\cdot\mathbb{E}\left[\mathrm{EE}[t]\mid \boldsymbol{\Theta}[t]\right], 
\end{align}
where $V > 0$ controls the trade-off between queue stability and EE maximization.
Applying the queue evolution and virtual-queue update equations, and upper-bounding the quadratic drift terms using standard Lyapunov techniques, we obtain the following upper bound for the drift-plus-penalty expression:
\begin{align}
  \Delta_V[t]
  \le
  C
  +
  \sum\limits_{k=1}^K
    \theta_k[t]
    \left(A_k[t] - \hat R_k[t]\right)
 - V\;\mathrm{EE}[t],
  \label{eq:drift_plus_penalty_bound}
\end{align}
where $C$ is a finite time-independent constant, $\mathbf{X}[t]$ denotes the binary clustering matrix in slot $t$, and $\theta_k[t]
  =
  Q_k[t]
  + \alpha_1 U_k[t]+ \alpha_2 Y_k[t]
  + \alpha_3 W_k[t]$
is a linear combination of its physical and virtual queues, which serves as a non-negative weight for UE $k$ in slot $t$ with $\alpha_1, \alpha_2, \alpha_3 \ge 0$.

Since the data arrivals ${A_k[t]}$ are exogenous and independent of the current queue states and clustering decisions, the summation $\sum_{k=1}^K \theta_k[t] A_k[t]$ is treated as an uncontrollable constant within time slot $t$. 
Therefore, after removing additive constants that do not affect the optimization, the objective of minimizing the upper bound in \eqref{eq:drift_plus_penalty_bound} reduces to solving the following per-slot deterministic optimization problem:
\begin{align} 
 \textbf{P2}:\ \mathop {{\rm{max}}}_{\mathbf{X}[t]} \  V \cdot \mathrm{EE}[t] + \sum\limits_{k=1}^K \theta_k[t]\,\hat R_k[t] 
\quad \text{s.t. C1}.
   \label{eq:P2_obj}
\end{align}
Problem~\textbf{P2} is a non-convex mixed-integer program (MINLP) characterized by: (i) binary variables $a_{k,l,m}[t] \in {0,1}$; (ii) a fractional objective function due to the instantaneous EE $\mathrm{EE}[t]$; and (iii) interference-coupled achievable rate $\hat R_k[t]$ within the objective.
To address this challenging MINLP, we develop an SCA-based approach, involving binary variable relaxation, the use of the quadratic transform for fractional terms, and the derivation of convex surrogates for non-convex components.

\subsection{Dynamic AP Clustering Algorithm}
\label{sec:sca_algorithm}
First, we relax the binary constraints on the association variables to $\widetilde{\mathrm{C1}}:\, 0 \le a_{k,l,m}[t] \le 1,\ \forall k,l,m$.
To promote binary solutions, we augment the objective with a penalty function $P(\mathbf{X}[t]) = \rho \sum_{k,l,m} \left( a_{k,l,m}[t] - a_{k,l,m}^2[t] \right)$, where $\rho > 0$ is a weighting parameter. 
This penalty term is zero if and only if $a_{k,l,m}[t] \in \left\{0,1\right\}$ for all $k,l,m$. Consequently, the relaxed per-slot optimization objective becomes
\begin{align}
 f\left(\mathbf{X}[t]\right)
\triangleq
V\cdot\mathrm{EE}[t]
+ \sum\limits_{k=1}^K \theta_k[t]\,\hat R_k[t]
- P\left(\mathbf{X}[t]\right).
  \label{eq:relaxed_objective}
\end{align}

\subsubsection{Auxiliary-Variable Reformulation via Quadratic Transform}
For any feasible clustering matrix $\mathbf{X}[t]$ yielding a non-negative sum rate $R_{\mathrm{sum}}[t] \ge 0$ and a positive total power consumption $P_{\mathrm{tot}}[t] > 0$, the weighted EE term $V \cdot \mathrm{EE}[t]$ can be recast using the quadratic transform for fractional programming~\cite{shen2018fractional}:
\begin{align}
\max_{y[t]\ge 0}\;
2y[t]\sqrt{V B\,R_{\mathrm{sum}}[t]}
- y^2[t]\,P_{\mathrm{tot}}[t].
\label{eq:quad_transform}
\end{align}
Given a fixed $\mathbf{X}[t]$, the optimal auxiliary variable is obtained in closed form as $y^\star[t] = \frac{\sqrt{V B\,R_{\mathrm{sum}}[t]}}{P_{\mathrm{tot}}[t]}$.
We therefore employ a two-block alternating optimization procedure within each time slot $t$: (i) update the auxiliary variable $y[t]$ according to the above closed-form expression given the current $\mathbf{X}[t]$;  (ii) with $y[t]$ fixed, solve for the clustering matrix $\mathbf{X}[t]$ by maximizing a convex surrogate constructed via SCA.

\subsubsection{Convex Surrogate via SCA}
With $y[t]$ held fixed, the transformed objective remains non-concave with respect to $\mathbf{X}[t]$, owing to the difference-of-convex (DC) structure of the penalty term and the interference coupling in the rate expressions. 
At SCA iteration $j$, we construct a concave surrogate by minimizing the non-concave components of the objective at the current iterate $\mathbf{X}^{(j)}[t]$. 
Specifically, we linearize the convex term $a_{k,l,m}^2[t]$ at $a_{k,l,m}^{(j)}[t]$ via its first-order Taylor approximation. 
Moreover, for each UE $k$, we replace $\hat R_k(\mathbf{X}[t])$ with a concave lower bound $\tilde R_k(\mathbf{X}[t];\mathbf{X}^{(j)}[t])$ that is tight at $\mathbf{X}^{(j)}[t]$ and satisfies
$\tilde R_k(\mathbf{X}[t];\mathbf{X}^{(j)}[t]) \le \hat R_k(\mathbf{X}[t])$ for all feasible $\mathbf{X}[t]$.
Substituting these approximations yields the overall concave surrogate objective for the $j$-th iteration:
\begin{align}
&\tilde f^{(j)}\!\left(\mathbf{X}[t];y[t]\right)
=
2y[t]\sqrt{V B\,\tilde R_{\mathrm{sum}}^{(j)}\!\left(\mathbf{X}[t]\right)}
- y^2[t]\,\tilde P_{\mathrm{tot}}^{(j)}\!\left(\mathbf{X}[t]\right) \nonumber\\
&\quad
+ \sum\nolimits_{k=1}^K \theta_k[t]\,
\tilde R_k^{(j)}\!\left(\mathbf{X}[t]\right)
- \tilde P^{(j)}\!\left(\mathbf{X}[t]\right).
\label{eq:sca_surrogate_short}
\end{align}
Thus, with $y[t]$ fixed, the next iterate $\mathbf{X}^{(j+1)}[t]$ is obtained by maximizing~\eqref{eq:sca_surrogate_short} over the relaxed constraints $\widetilde{\mathrm{C1}}$. 
The inner SCA loop terminates when the relative improvement in the surrogate objective falls below a predefined tolerance $\epsilon_{\mathrm{SCA}} > 0$, or when a preset maximum number of iterations $J_{\max}$ is reached.
Finally, the continuous-valued solution $\mathbf{X}^\star[t]$ from the SCA procedure is projected back to the binary domain $\{0,1\}$ to obtain a feasible AP clustering decision.

\section{Simulation Results}
\label{sec:sim_results}

We consider a CF-RAN over a $500\times500~\mathrm{m}^2$ wrap-around area.
The network consists of $M=2$ EDUs coordinating $L=20$ distributed APs, each equipped with $N=6$ antennas, to serve $K=8$ UEs.
AP and UE locations are randomly generated.
The system bandwidth is $B=20~\mathrm{MHz}$ and the coherence block length is $\tau_{\rm{c}}=200$ symbols with $\tau_{\rm{p}}=2$ pilot symbols.
We use UL pilot power $\rho_p=0.2~\mathrm{W}$ and a DL power budget parameter $\rho_{\mathrm{d}}=1~\mathrm{W}$.
For queueing control, we set $A_k^{\max}=2~\mathrm{bits/slot}$, $Q_0=1.5$, $\varepsilon_k^{(Q)}=0.01$, and $V=5$.
Our power model parameters follow those specified in~\cite{ngo2018total_ee_cf}. 
For fair comparison, the baseline is a queue-aware but tail-unaware user-centric clustering policy that optimizes AP-UE associations using queue-state information and large-scale channel conditions, without exploiting EVT-based tail feedback.
After association is fixed, the same equal-power FZF precoding is applied to both schemes.
\begin{figure}[t!]
  \centering
  \subfloat[Exceedance process]{%
    \begin{minipage}[t]{0.95\columnwidth}
      \centering
      \includegraphics[width=1\linewidth]{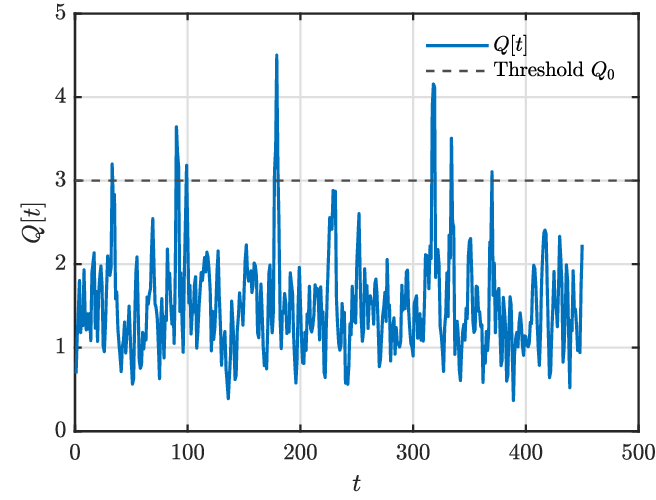}%
      \label{subfig:evt_method_a}
    \end{minipage}
  }%
  \hspace{3pt}%
  \subfloat[Second-order moment]{%
    \begin{minipage}[t]{0.95\columnwidth}
      \centering
      \includegraphics[width=1\linewidth,height=0.792\linewidth]{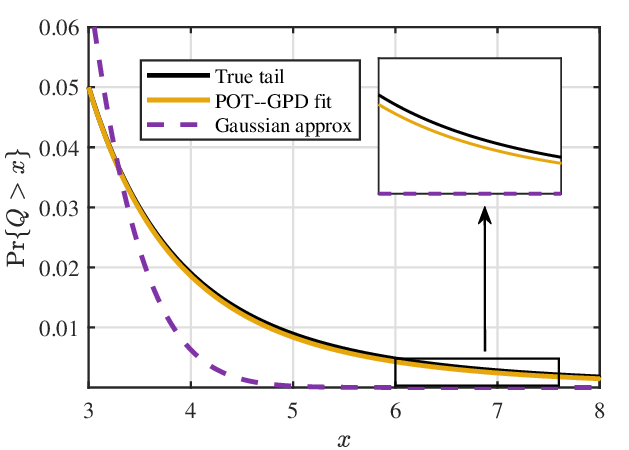}%
      \label{subfig:evt_method_b}
    \end{minipage}
  }%
  \caption{Illustrative POT-GPD validation and tail-moment behavior for the exceedance process.}
  \label{fig:delay_second_moment}
\end{figure}


\begin{figure}
	\centering
		\includegraphics[width=0.95\linewidth]{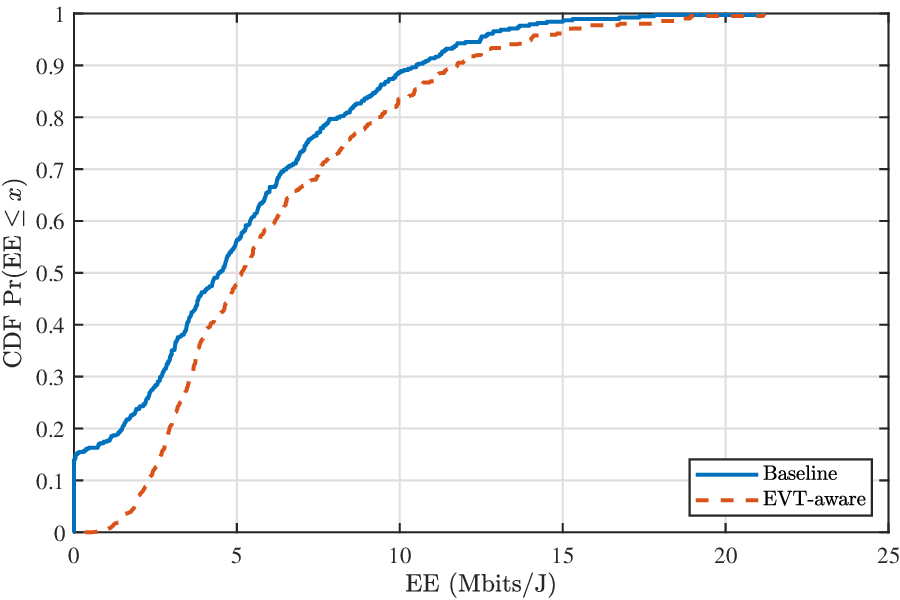}
		\caption{Empirical CDF of steady-state EE.}
		\label{fig:ss_ee_cdf}
\end{figure}

\begin{figure}
		\centering  		
        \includegraphics[width=0.95\linewidth]{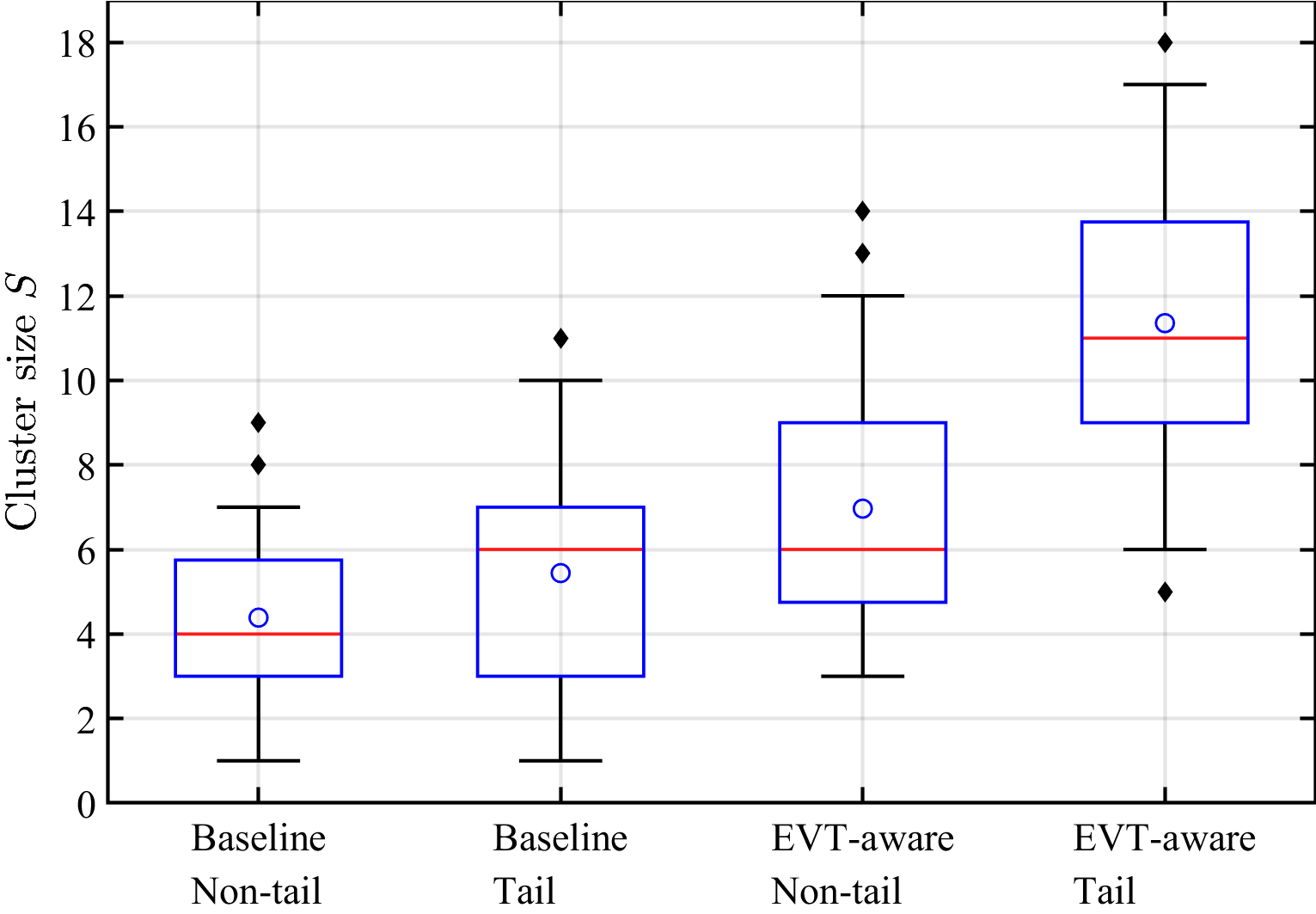}
		\caption{Dynamic cluster-size distribution.}
		\label{fig:cluster_size_box}
\end{figure}

Fig.~\ref{fig:delay_second_moment} provides a conceptual and empirical illustration of the POT-GPD modeling framework used in this study. 
Fig.~\ref{fig:delay_second_moment}(a) illustrates the exceedance process, where samples above the threshold $Q_0$ constitute the POT dataset. 
Fig.~\ref{fig:delay_second_moment}(b) shows that the POT-GPD model provides a markedly tighter fit in the high-exceedance region than a Gaussian baseline.

\cref{fig:ss_ee_cdf} plots the empirical cumulative distribution function (CDF) of the steady-state EE. 
The EVT-aware scheme exhibits a right-shifted CDF, indicating a higher probability of operating at larger EE and a reduced likelihood of low-EE realizations. 
This gain comes from queue-tail-aware clustering, which adapts the cooperation set to congestion states and uses AP cooperation more efficiently than the queue-unaware baseline under the same equal-power FZF transmission.

To further reveal how EVT-aware clustering reacts to critical queue buildups, we examine the conditional distribution of the hard cluster size $S_k[t]$ under tail and non-tail regimes.
As shown in \cref{fig:cluster_size_box}, the baseline scheme maintains relatively small clusters with limited sensitivity to the queue state, leading to similar cluster-size distributions across the two regimes.
In contrast, the EVT-aware scheme selects substantially larger clusters on average.
Specifically, conditioned on a tail event ($Q_k[t]\ge Q_0$), the cluster-size distribution shifts upward compared with the non-tail regime, indicating event-driven resource pooling for highly congested UEs.
This tail-triggered increase in cooperation improves the effective SINR and yields higher service rates, thereby mitigating queue backlogs.

\section{Conclusion}
This paper proposes a user-centric clustering framework for CF-RAN that addresses the joint challenges of FBL transmission and queue-aware reliability.
The framework characterizes rare but critical queueing events through a POT–GPD approximation, leading to a set of tractable and interpretable tail-risk metrics. 
Building on these metrics, an online optimization framework is developed based on Lyapunov drift-plus-penalty theory. 
The resulting per-slot problem is efficiently solved using an SCA approach, which employs binary variable relaxation and a penalty method.
Simulation results show that the proposed EVT-aware clustering strategy effectively suppresses extreme queueing delays while maintaining competitive EE.
This yields a superior reliability-efficiency trade-off, crucial for meeting the stringent demands of uRLLC services.








\bibliographystyle{IEEEtran}
\bibliography{IEEEabrv,mybibfile}









\end{document}